\documentclass[a4paper,11pt]{article}

\usepackage{jheppub} % for details on the use of the package, please see the JINST-author-manual
\usepackage{bm,latexsym,amsmath,amsfonts,amssymb,fancyhdr,color,graphicx,slashed}
\usepackage[usenames,dvipsnames,svgnames,table]{xcolor}
\usepackage[normalem]{ulem}
\usepackage{soul}

\setstcolor{red}

\makeatletter
\global\def\@fpheader{\,}
\makeatother

\graphicspath{{figs/}} % set the path of figures

\title{\boldmath Boosted Dark Matter From Centaurus A \\ and Its Detection}

\author[b,c]{Chen Xia,}
\author[b,c]{Chuan-Yang Xing,}
\author[a]{Yan-Hao Xu}
\affiliation[a]{Department of Physics, Yantai University, Yantai 264005, China}
\affiliation[b]{Tsung-Dao Lee Institute \& School of Physics and Astronomy, Shanghai Jiao Tong University, China}
\affiliation[c]{Key Laboratory for Particle Astrophysics and Cosmology (MOE) \& Shanghai Key Laboratory for Particle Physics and Cosmology, Shanghai Jiao Tong University, Shanghai 200240, China}

\emailAdd{xiachen@sjtu.edu.cn}
\emailAdd{chuan-yang-xing@sjtu.edu.cn}
\emailAdd{xuyanhao@ytu.edu.cn}

\abstract{
    Dark matter can be boosted by high energy particles in astrophysical environments through elastic scattering. 
    We study the production of boosted dark matter via scattering with electrons in the relativistic jet of the closest active galactic nucleus, Centaurus A, and its detection in the Super-Kamiokande experiment.
    Since there are a huge number of electrons in the jet and dark matter is extremely dense around the supermassive black hole that powers the jet, the number of boosted dark matter is tremendously large.
    Compared to boosted dark matter from blazars, the dark matter flux from Centaurus A is enhanced due to the proximity of Centaurus A. 
    The constraint on dark matter-electron scattering cross section set by Super-Kamiokande is more stringent, down to $\sim 10^{-36} \, \mathrm{cm}^2$ for $\mathrm{MeV}$ dark matter.
    }
\begin{document}
\maketitle
\flushbottom

%%%%%%%%%%%%%%%%%%%%%%%%%%%%%%%%%%%%%%%%%%%%%%%%%%%%%%%%%%%%%%%%
\section{Introduction}\label{sec:intro}
%%%%%%%%%%%%%%%%%%%%%%%%%%%%%%%%%%%%%%%%%%%%%%%%%%%%%%%%%%%%%%%%

With enormous cosmological and astrophysical observations suggesting the existence of dark matter (DM), the nature of DM remains a mystery~\cite{Bertone:2004pz,Bauer:2017qwy,Cooley:2021rws}.
DM in the form of exotic particles is well motivated by theories, notably the Weakly Interacting Massive Particles (WIMPs) paradigm, which predicts DM particles with mass and coupling strength similar to the electroweak sector of the Standard Model (SM).
Inspired by WIMPs, numerous DM direct detection experiments~\cite{SENSEI:2019ibb,PandaX-4T:2021bab,CDEX:2022fig,LZ:2022lsv,SuperCDMS:2020aus,DarkSide-50:2022qzh,XENON:2023cxc} have been established and improved in the past decades.
However, still no conclusive DM signal has been detected.
The constraints imposed by direct detection experiments on WIMPs have become increasingly stringent.

In the past years, sub-GeV mass scale DM has attracted increasing attention~\cite{SENSEI:2019ibb,XENON:2019gfn,SuperCDMS:2020aus,PandaX-II:2021kai,PROSPECT:2021awi,COHERENT:2021pvd,CDEX:2022kcd,Super-Kamiokande:2022ncz,CRESST:2022lqw,DAMIC-M:2023gxo,Knapen:2017xzo,Evans:2019jcs,Elor:2021swj}. 
Inside conventional direct detection experiments, sub-GeV halo DM particles are too light to trigger detectable signals  and are much less constrained. 
To enhance the detectability of sub-GeV DM in the direct detection experiments, various kinds of boosted DM scenarios have been proposed, such as DM boosted by annihilation of heavy DM~\cite{Agashe:2014yua}, solar reflection~\cite{An:2017ojc,Emken:2021lgc,An:2021qdl}, high-energy cosmic rays in the Milky Way~\cite{Bringmann:2018cvk,Ema:2018bih,Cappiello:2019qsw,Wang:2019jtk,Ge:2020yuf,Guo:2020drq,Xia:2020apm,Cao:2020bwd,Xia:2021vbz,Maity:2022exk,Wang:2023wrx,Su:2023zgr},
cosmic ray-atmosphere inelastic collision~\cite{Alvey:2019zaa,Flambaum:2020xxo},
and diffuse supernova neutrino background~\cite{DeRomeri:2023ytt}.
The boosted DM carries enough energy to deposit recoil energy above the threshold in DM direct detection experiments~\cite{PandaX-II:2021kai,CDEX:2022fig} or neutrino experiments~\cite{Super-Kamiokande:2022ncz,PROSPECT:2021awi}.

Recently, the scenario of blazar boosted DM has shown the potential to derive more stringent constraints~\cite{Wang:2021jic,Granelli:2022ysi,Bhowmick:2022zkj}.
Blazar is a type of active galactic nucleus (AGN) with a relativistic jet directed very nearly towards the Earth.
AGN is theorized to contain a supermassive black hole in the center which produces the jet.
The relativistic particles in the jet can boost DM particles around the supermassive black hole through leptonic or hadronic scattering.
The boosted DM propagating to the Earth can be detected in underground experiments.

In this work, we propose that the boosted DM from the closest AGN, the AGN in the galaxy Centaurus A (Cen A), can dominate over that from blazars studied in Refs.~\cite{Wang:2021jic,Granelli:2022ysi} 
assuming DM is coupled to electrons.
Note that Cen A is not a blazar since its jets do not point towards the Earth. 
Instead, the viewing angle between the jet axis and our line-of-sight to Cen A is quite large, which is measured to be $ \theta_\mathrm{LOS} \sim 12^\circ \mathrm{-} 45^\circ$~\cite{Muller:2014wja}.
Inside the jet of Cen A, the majority of charged particles propagate along the jet axis.
Since the particles in the jet are highly relativistic, the upscattered light DM is primarily forward. 
Therefore, the boosted DM particles mostly travel along the jet as well.
The number of boosted DM particles with a large collimation angle, $\sim \theta_{\mathrm{LOS}}$ with respect to the jet axis, so that they can travel to the Earth, is suppressed. 
That is, the fraction of boosted DM from Cen A to the Earth is lower than the fraction from blazars.
However, we note that Cen A is much closer to us than blazars. 
The luminosity distance of Cen A is $d_L \simeq 3.7 \, \mathrm{Mpc}$~\cite{Harris:2009wj,Banik:2020ffo}. 
For comparison, the luminosity distance of the blazar BL Lacertae (BL Lac) studied in Refs.~\cite{Wang:2021jic,Granelli:2022ysi} is $\simeq 322.7 \, \mathrm{Mpc}$ around two orders of magnitude further than Cen A. 
Therefore, we can expect a four-order-of-magnitude enhancement of DM flux from Cen A at the Earth than that from BL Lac.
This can compensate for the suppression of boosted DM moving at large angles within the jet. 
The astrophysical environments of Cen A and BL Lac are similar~\cite{Boettcher:2013wxa,Banik:2020ffo}, so we can expect the total amounts of boosted DM produced in these two AGNs to be of the same order.
Consequently, with the suppression from viewing angles and the enhancement from luminosity distances taken into account, the flux of boosted DM from Cen A is about two orders of magnitude higher than that from blazars.
The constraints on the DM-electron scattering cross section from the Super-Kamiokande (Super-K) experiment can be improved by around one order of magnitude.
In addition, the Earth attenuation effect on DM particles before reaching the underground detector is considered in this work.
This effect obstructs DM particles with large cross sections, leading to an upper boundary of the exclusion region.

This paper is organized as follows: In section~\ref{sec:jet}, we introduce the distribution of electrons inside the jet of Cen A.
In section~\ref{sec:flux}, we calculate the flux of boosted DM. 
The attenuation of boosted DM inside the Earth is carried out in section~\ref{sec:attenuation}.
Then, we analyze the detection of the boosted DM in Super-K detector and derive constraints in section~\ref{sec:detection}.
We summarize our results in section~\ref{sec:summary}.

%%%%%%%%%%%%%%%%%%%%%%%%%%%%%%%%%%%%%%%%%%%%%%%%%%%%%%%%%%%%%%%%
\section{The Relativistic Jet of Centaurus A}\label{sec:jet}
%%%%%%%%%%%%%%%%%%%%%%%%%%%%%%%%%%%%%%%%%%%%%%%%%%%%%%%%%%%%%%%%

The distribution of high-energy electrons in the relativistic jet is an essential ingredient in the production of boosted DM.
Our understanding of the distribution of electrons is based on electromagnetic observations of the core region of Centaurus A (Cen A). 
The observations span a range from microwaves to $\gamma$-rays, and are typically plotted in energy versus wavelength, known as the spectral energy distribution (SED). 
Similar to other AGNs, the SED of Cen A exhibits two peaks (see in e.g. Ref.~\cite{Banik:2020ffo}): 
the lower-energy one is located in the infrared band $(\simeq 1 \, \mathrm{eV})$, and the higher-energy peak is in the $\gamma$-ray region $(\simeq 1 \, \mathrm{MeV})$.
These two peaks are theorized to have different physical origins~\cite{Ghisellini:1998it,Moderksi:2005jw,Boettcher:2006pd,Celotti:2007rb,Finke:2008pe,Boettcher:2013wxa,Blandford:2018iot,Cerruti:2020lfj}. 
The first peak is dominated by the synchrotron radiation of relativistic charged particles within the jet, where strong tangled magnetic fields exist. 
On the other hand, the second peak is due to the inverse Compton scattering of high-energy particles with background photons in the jet.

Based on these two interactions, there can be various models to reproduce the two-peak SEDs.
Firstly, the dominant types of charged particles can be different. 
In a leptonic model, the electromagnetic radiation is mainly produced by electrons, whereas in a hadronic model, it results dominantly from protons. 
In addition, the number of emission regions within the jet can vary.

Here we employ the simple two-zone model from Ref.~\cite{Banik:2020ffo} for the relativistic jet of Cen A. 
In this model, one of the emission regions locates in the inner jet that is close to the supermassive black hole at the center of Cen A, called the core emission region. 
The other emission region is considerably distant $(\sim \mathrm{kpc})$ from the inner jet. 
Since DM is dense in the region near the supermassive black hole (see Sec.~\ref{sec:flux}), we can concentrate on the distributions of charged particles in the core emission region and neglect the $\mathrm{kpc}$ emission region. 
In the core emission region, the dominant charged particles are electrons. 

The core emission region is modeled as a spherical blob with radius $R'_b$. 
Note that in this paper, we use primed symbols to denote quantities in the blob's comoving frame and use non-primed ones for those in the observer's frame.
The blob propagates outwards along the jet relativistically with a boost factor $\Gamma_B$ and velocity $\beta_B = \sqrt{1- \Gamma_B^{-2}}$. 
In the blob's comoving frame, electrons are homogeneously distributed within the blob and move isotropically. 
The spectrum of the electrons is assumed to follow a broken power-law~\cite{Banik:2020ffo},
\begin{equation}
    \frac{dn'_e}{dE'_e}
    = 
    \begin{cases}
        k_e \left( \frac{E'_e}{m_e} \right)^{-\alpha_1}, & \gamma'_{e,{\min}} \leq E'_e/m_e \leq \gamma'_{b}  \\
        k_e (\gamma'_{b})^{\alpha_2 - \alpha_1} \left( \frac{E'_e}{m_e} \right)^{-\alpha_2}, & \gamma'_{b} \leq E'_e/ m_e \leq \gamma'_{e,{\max}}
    \end{cases},
    \label{electron_distribution_blob_frame}
\end{equation}
where $\gamma'_{e,{\min}}$ and $\gamma'_{e,{\max}}$ denote the minimal and maximal boost factors of electrons in the blob, respectively, and $\gamma'_{b}$ represents the break boost factor of the spectrum.
The normalization constant $k_e$ can be determined by the comoving luminosity of electrons, $L'_e$, in the blob's comoving frame~\cite{Banik:2019jlm},
\begin{equation}
    L'_e 
    = 
    \pi R^{\prime 2}_B \beta_B 
    \int_{\gamma'_{e,{\min}} m_e}^{\gamma'_{e,{\max}} m_e} 
    E'_e \frac{dn'_e}{dE'_e}
    dE'_e .
\end{equation}

The distribution of electrons in the observer's frame is connected with Eq.~\eqref{electron_distribution_blob_frame} by a Lorentz boost,
\begin{equation}
    \frac{d^2n_{e}}{dE_e d\Omega_e} 
    = 
    \Gamma_B 
    \left|
        \begin{array}[2]{cc}
            \frac{\partial E'_e}{\partial E_e} & \frac{\partial E'_e}{\partial u} \\
            \frac{\partial u'}{\partial E_e} & \frac{\partial u'}{\partial u} 
        \end{array}
    \right|
    \frac{d^2n'_{e}}{dE'_e d\Omega'_e} .
\end{equation}
Here $u \equiv \cos \theta_e$ ($u' \equiv \cos \theta'_e$) represents the cosine of the angle between the electron momentum and the jet axis in the observer's (blob's comoving) frame.
The factor $\Gamma_B$ arises from the length contraction effect of relativistically moving objects.
Using the assumption that the electrons are isotropically distributed in the blob frame, $\frac{d^2n'_{e}}{dE'_e d\Omega'_e} = \frac{1}{4\pi} \frac{dn'_{e}}{dE'_e}$, we can calculate the explicit expression of the electron distribution in the observer's frame as follows,
\begin{equation}
    \frac{d^2n_{e}}{dE_e d\Omega_e} 
    = 
    \frac{1}{4\pi}
    \frac{ \beta_e }{ \sqrt{(1 - \beta_B \beta_e u)^2 - (1-\beta_B^2)(1-\beta_e^{2}) } }
    \frac{dn'_e}{dE'_e}(E'_e) ,
    \label{electron_distribution_obs_frame}
\end{equation}
where $\beta_e = \sqrt{1-\left(\frac{E_e}{m_e}\right)^{-2}}$ is the velocity of the electron and $E'_e  = \Gamma_B ( 1 - \beta_B \beta_e u ) E_e$.

Note that the radius of the blob is smaller than the length of the inner jet.
In this simple blob model, electrons are predominantly distributed inside the blob, and the number of electrons outside the blob is negligible. 
Therefore, in addition to the distribution of electrons in Eq.~\eqref{electron_distribution_obs_frame}, we also need to know the position of the blob to calculate the boosted DM flux, since the number density of DM particles varies at different locations. 
We assume the complete picture for the jet is that the blob is propagating outwards within the inner jet.
When a blob reaches the boundary of the inner jet, it has lost most of its energy and a new blob is born near the supermassive black hole.
In this picture, there is only one blob at a time.
The inner jet of Cen A has been imaged by the Event Horizon Telescope~\cite{EventHorizonTelescope:2021iqj} with high resolution.
According to the imaged jet structure, the length of the core emission region is approximately $ L_\mathrm{core} \simeq 500 R_\mathrm{S} $, where $R_\mathrm{S}$ denotes the Schwarzschild radius of the supermassive black hole.
Since the speed of the blob is close to the speed of light, the lifetime of a blob can be estimated as $\sim 10^5 \, \mathrm{s}$.
Therefore, to calculate the number of boosted DM over a period as long as years, we can effectively use the average number density within the whole inner jet, which is given by
\begin{equation}
    \frac{d^2n_{e, \mathrm{eff}}}{dE_e d\Omega_e}
    =
    \frac{R_B}{L_\mathrm{core}} 
    \frac{d^2n_{e}}{dE_e d\Omega_e}
    =
    \frac{R'_B / \Gamma_B}{L_\mathrm{core}} 
    \frac{d^2n_{e}}{dE_e d\Omega_e} ,
    \label{eq:effective_electron_distribution}
\end{equation}
Note that the averaging factor $R_B/L_\text{core}$ is not considered in the previous work on blazar boosted DM~\cite{Granelli:2022ysi}, so our calculation is more conservative.

\begin{figure}[t]
    \centering
    \includegraphics[width=0.7\textwidth]{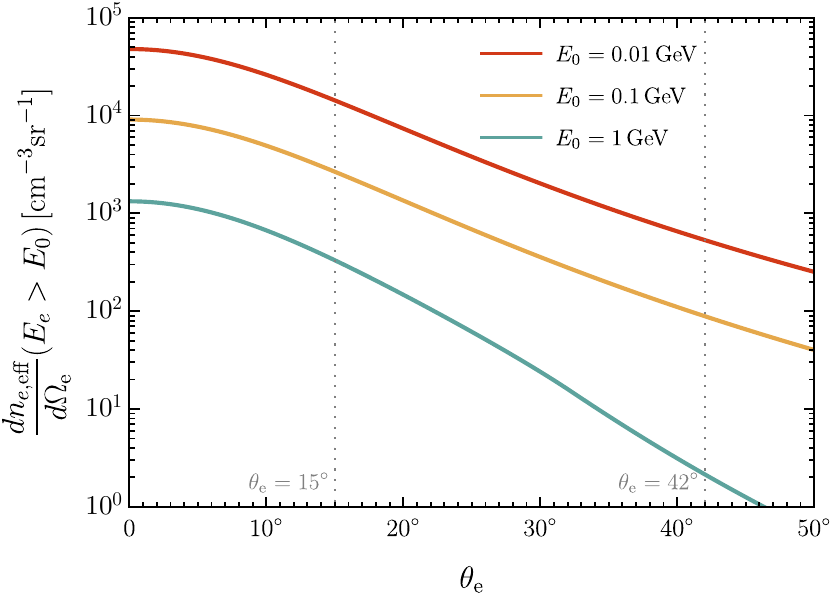}
    \caption{Effective distribution of electrons concerning the incident angle $\theta_e$ with energy larger than $E_0$.
    The minimum $E_0$ is taken to be $0.01 \, \mathrm{GeV}$, $0.1 \, \mathrm{GeV}$, and $1 \, \mathrm{GeV}$ for the red, yellow, and green curve, respectively.
    }
    \label{fig:electron}
\end{figure}

Note again that the relativistic jet of Cen A is not directed towards the Earth.
To produce boosted DM propagating to the Earth, the oblique angle between the momentum of boosted DM and the jet axis should be $\theta_\chi = \theta_{\mathrm{LOS}}$, where $\theta_{\mathrm{LOS}}$ is the viewing angle between the jet axis and our line-of-sight to Cen A.
This angle $\theta_\chi$ is determined by the incident angle $\theta_e$ of the initial state electron and the scattering angle.
To get a large $\theta_\chi$, we need a large incident angle $\theta_e$ or a large scattering angle. 
Here we firstly investigate the distribution of electrons with respect to the incident angle $\theta_e$ between the electron momentum and the jet axis.
For this purpose, we define the effective distribution of electrons with energy exceeding a minimum $E_0$ as,
\begin{equation}
    \frac{d n_{e, \mathrm{eff}}}{ d\Omega_e} ( E_e > E_0 )
    = 
    \int_{E_0} \frac{d^2n_{e, \mathrm{eff}}}{dE_e d\Omega_e} dE_e.
\end{equation}
According to Eq.~\eqref{eq:effective_electron_distribution} and the parameters about the jet of Cen A adopted from Ref.~\cite{Banik:2020ffo} (see Tab.~\ref{tab:parameters} below), in Fig.~\ref{fig:electron}, we show the effective distribution versus the incident angle $\theta_e$ with three different minima $E_0$. 
We can see that most electrons are propagating along the jet axis.
This is because the Lorentz boost from the blob's comoving frame to the observer's frame is in this direction. 
We note that the portion of electrons traveling at a relatively large angle against the jet axis is non-negligible.
For example, more than one-third of electrons have an incident angle $\theta_e > 15^\circ$, and around one percent of electrons propagate with $\theta_e > 42^\circ$.
These electrons with large incident angles can boost DM particles with large oblique angles, allowing them to travel to the Earth.
Therefore, compared to the scenario in which DM is boosted in blazars~\cite{Wang:2021jic,Granelli:2022ysi}, the fraction of boosted DM propagating towards the Earth from Cen A is suppressed by one to two orders of magnitude at most due to the large viewing angle.

%%%%%%%%%%%%%%%%%%%%%%%%%%%%%%%%%%%%%%%%%%%%%%%%%%%%%%%%%%%%%%%%
\section{Boosted Dark Matter Flux}\label{sec:flux}
%%%%%%%%%%%%%%%%%%%%%%%%%%%%%%%%%%%%%%%%%%%%%%%%%%%%%%%%%%%%%%%%

Another essential ingredient for calculating boosted DM flux is the distribution of DM particles in Cen A.
Around supermassive black holes that grow adiabatically, DM is accreted to form a dense spike~\cite{Gondolo:1999ef}. 
Consider a DM halo model with an inner cusp before the black hole formation, and the initial density profile follows a simple power-law, $\rho (r) \propto r^{-\gamma}$, where $\gamma$ is the power-law index. 
As the formation of the supermassive black hole, the DM profile will evolve to be~\cite{Gondolo:1999ef},
\begin{equation}
    \rho'(r)
    \propto 
    \left( 1- \frac{4R_\mathrm{S}}{r} \right)^3 
    r^{- \gamma_\mathrm{sp}} ,
    \qquad
    \mathrm{with}
    \qquad
    \gamma_{\mathrm{sp}}
    = 
    \frac{9-2\gamma}{4-\gamma} .
    \label{spike_profile_propto}
\end{equation}
For DM halos initially following the Navarro–Frenk–White (NFW) profile~\cite{Navarro:1995iw}, $\gamma = 1$, we have $\gamma_\mathrm{sp} = \frac{7}{3}$. 
That is, the final DM profile is much steeper than the initial one. 
If DM particles are too close to the supermassive black hole, they will fall into the black hole.
The DM number density vanishes at a radius $r < 4R_\mathrm{S}$~\cite{Gondolo:1999ef}. 
The normalization constant for the profile $\rho'(r)$ can be determined by the DM mass enclosed within a large radius. 
Following Ref.~\cite{Ullio:2001fb,Gorchtein:2010xa,Wang:2021jic,Granelli:2022ysi,Cermeno:2022rni,Herrera:2023nww}, the mass of enclosed DM can be of the same order as the black hole mass,
\begin{equation}
    \int_{4 R_\mathrm{S}}^{{10^5 R_\mathrm{S}}}
    4 \pi r^2 
    \rho'(r) 
    dr 
    \simeq 
    M_{\mathrm{BH}} .
\end{equation}
In this normalization relation, the upper limit is conventionally taken to be $10^5 R_\mathrm{S}$~\cite{Gorchtein:2010xa,Wang:2021jic,Granelli:2022ysi}.

If DM can annihilate with each other, the annihilation channel will reduce the DM density in the spike, setting a maximal DM energy density to be $\rho_\mathrm{core} = m_\chi / \left\langle \sigma v_\mathrm{rel} \right\rangle t_\mathrm{BH}$~\cite{Gondolo:1999ef}.
Here $m_\chi$ denotes the DM particle mass; $\left\langle \sigma v_\mathrm{rel} \right\rangle$ is the thermal averaging of its annihilation cross section times relative velocity; $t_\mathrm{BH}$ is the lifetime of the black hole.
The final DM spike profile becomes~\cite{Gondolo:1999ef}
\begin{equation}
    \rho_{\mathrm{sp}} (r)
    = 
    \frac{\rho'(r) \rho_{\mathrm{core}} }{ \rho'(r) + \rho_{\mathrm{core}} } .
    \label{spike_profile}
\end{equation}

With the electron distribution in Eq.~\eqref{eq:effective_electron_distribution} and the DM profile in Eq.~\eqref{spike_profile}, we can calculate the production rate of boosted DM inside the jet. 
The number of boosted DM particles produced per unit volume per unit time can be written as
\begin{equation}
    \frac{dn_\chi}{dt} 
    = 
    \int 
    \frac{d^2n_{e, \mathrm{eff}}}{dE_e d\Omega_e} 
    n_\mathrm{sp}
    \frac{d\sigma_{\chi e}}{d u_\mathrm{sc}}
    \beta_e
    dE_e d\Omega_e du_\mathrm{sc}.
\end{equation}
In this equation, $n_\mathrm{sp} = \frac{\rho_\mathrm{sp}}{m_\chi}$ is the DM number density. 
$\frac{d\sigma_{\chi e}}{d u_\mathrm{sc}}$ denotes the differential cross section with respect to the cosine of the scattering angle in the observer's frame, $u_\mathrm{sc} \equiv \cos \theta_{\mathrm{sc}}$.
The non-relativistic motion of DM in the spike is neglected, and the relative velocity of DM and electrons is equal to the electron velocity $\beta_{e} \approx 1$.
The total number of boosted DM particles is obtained by integrating $\frac{dn_\chi}{dt}$ over the jet volume, $\frac{dN_\chi}{dt} = \pi R_B^{\prime 2} \int \frac{dn_\chi}{dt} dr$.
We are interested in the spectrum of boosted DM particles travelling towards the Earth, rather than the total number of boosted DM particles.
To achieve this, we can introduce two $\delta$-functions into the expression for $\frac{dN_\chi}{dt}$,
\begin{equation}
    \frac{dN_\chi}{dt dE_\chi d\Omega_\chi} 
    = 
    \frac{R_B^{\prime 2} }{2} 
    \int 
    \frac{d^2n_{e, \mathrm{eff}}}{dE_e d\Omega_e} 
    n_\mathrm{sp}
    \frac{d\sigma_{\chi e}}{d u_\mathrm{sc}}
    \beta_e 
    \delta (E_\chi - E_\chi (E_e, u_\mathrm{sc}) )
    \delta^2 (\Omega_\chi - \Omega_\chi(\Omega_e, \Omega_\mathrm{sc}) )
    dE_e d\Omega_e d\Omega_\mathrm{sc} dr.
    \label{boost_rate_new_1}
\end{equation}
Here $E_\chi$ is the energy of the boosted DM, and $\Omega_\chi$ is the solid angle of the boosted DM momentum. 
$\Omega_\mathrm{sc}$ is the scattering solid angle of the outgoing DM momentum $\mathbf{p}_\chi$ relative to the injecting electron momentum $\mathbf{p}_e$.
The function $\Omega_\chi(\Omega_e, \Omega_\mathrm{sc})$ indicates that $\Omega_\chi$ can be fully determined by the solid angle of the injecting electron $\Omega_e$ and the scattering solid angle $\Omega_\mathrm{sc}$.
Furthermore, the function $E_\chi (E_e, u_\mathrm{sc})$ describes the relation between the energy of the boosted DM and the injecting electron energy and the scattering angle.
Its form can be determined by the energy-momentum conservation condition of the process,
\begin{equation}
    E_\chi( E_e, u_\mathrm{sc} ) 
    = 
    \frac{ \gamma_\mathrm{CM}^2 + (\gamma_\mathrm{CM}^2 - 1) u_\mathrm{sc}^2 }{ \gamma_\mathrm{CM}^2 - (\gamma_\mathrm{CM}^2 - 1) u_\mathrm{sc}^2 } m_\chi ,
\end{equation}
where $\gamma_\mathrm{CM} = \frac{E_e + m_\chi}{ \sqrt{(E_e + m_\chi)^2 - |\mathbf{p}_e|^2} }$ is the boost factor from the center-of-mass frame to the observer's frame.

\begin{figure}[t]
    \centering
    \includegraphics[width=0.7\textwidth]{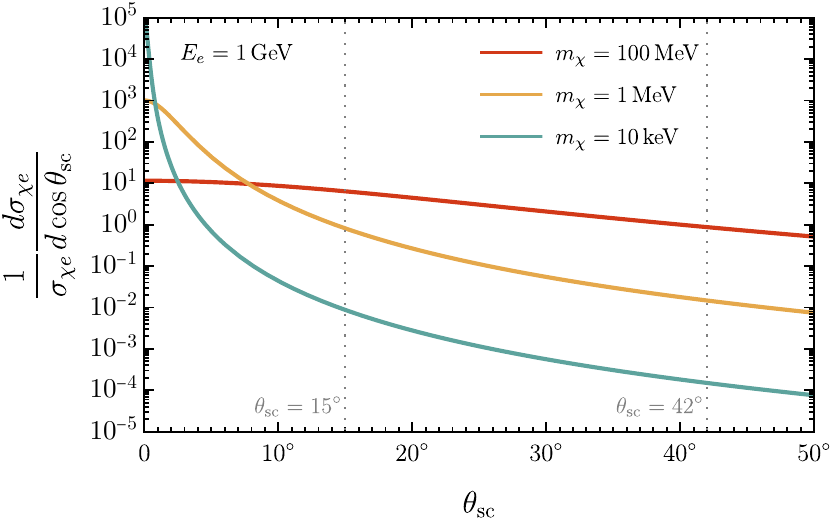}
    \caption{Differential cross sections versus scattering angles in the observer's frame.
    The injecting electron energy is set to be $E_e = 1  \, \mathrm{GeV}$. 
    The red, yellow, and green curves correspond to different DM masses, $m_\chi = 100 \, \mathrm{MeV}, \, 1 \, \mathrm{MeV}$, and $10 \, \mathrm{keV}$, respectively.
    }
    \label{fig:scattering}
\end{figure}

As discussed in Sec.~\ref{sec:jet}, to produce a boosted DM particle with a large oblique angle against the jet axis, we need a large incident angle $\theta_e$ or a large scattering angle $\theta_\mathrm{sc}$. 
The distribution of scattering angle is described by the differential cross section $\frac{d\sigma_{\chi e}}{d u_\mathrm{sc}}$.
We assume that the scattering of DM and electrons is isotropic in the center-of-mass frame. 
The differential cross section in the observer's frame is then,
\begin{equation}
    \frac{d\sigma_{\chi e}}{d u_\mathrm{sc}}
    = 
    \frac{4 \gamma_\mathrm{CM}^2 u_\mathrm{sc}}{ 
    \left( \gamma_\mathrm{CM}^2 - (\gamma_\mathrm{CM}^2 - 1) u_\mathrm{sc}^2 \right)^2 } 
    \frac{\sigma_{\chi e}}{2} .
    \label{dif_cross_section}
\end{equation}
Note that the $\theta_\mathrm{sc}$ in $u_\mathrm{sc}$ is the scattering angle of the upscattered DM.
In Fig.~\ref{fig:scattering}, we show the differential cross section for different DM masses with $E_e = 1 \, \mathrm{GeV}$.
We can see that the scattering is mainly forward in the observer's frame. 
Especially for light DM with $m_\chi \lesssim 1 \, \mathrm{MeV}$, the possibility of scattering processes with a scattering angle $\theta_\mathrm{sc} > 15^\circ$ is significantly suppressed. 
This means that to produce boosted light DM particles travelling towards the Earth, we need the incident angle of electrons to be large in the first place, $\theta_e \approx \theta_\mathrm{LOS}$.
For heavier DM with $m_\chi \simeq 100 \, \mathrm{MeV}$, the suppression from large scattering angle is modest, even smaller than the suppression arising from the electron number densities with large incident angles as shown in Fig.~\ref{fig:electron}.
In this case, the dominant contribution to boosted DM  comes from the scattering of the electrons moving along the jet with large scattering angles.

In the case of isotropic scattering of DM and electrons, we can integrate out the $\delta$-functions in Eq.~\eqref{boost_rate_new_1} to obtain,
\begin{equation}
    \frac{dN_\chi}{dt dE_\chi d\Omega_\chi} 
    = 
    \pi R_B^{\prime 2} 
    \sigma_{\chi e} 
    \int \frac{\rho_\mathrm{sp}}{m_\chi} dr 
    \times 
    \int_{E_e^{\min}}
    \frac{dE_e}{\Delta E_\chi^{\max}} 
    \frac{d \varphi_{\mathrm{sc}}}{2\pi} 
    \frac{dn_{e, \mathrm{eff}}}{dE_e d\Omega_e} 
    \beta_e ,
\label{eq:production_rate}
\end{equation}
where $\Delta E_\chi^{\max} = 2 ( \gamma_\mathrm{CM}^2 - 1) m_\chi$ is the maximal energy transfer from the electron to the DM, and the integral lower limit $E_e^{\min}$ is the minimal electron energy producing boosted DM with energy $E_\chi$,
\begin{equation}
    E_e^{\min} 
    = 
    \frac{E_\chi - m_\chi}{2}
    \left(
    1 + 
    \sqrt{
    \frac{(E_\chi + m_\chi)(m_\chi (E_\chi - m_\chi)+ 2 m_e^2 )}{m_\chi (E_\chi - m_\chi)^2}
    }
    \right) .
\label{eq:Eemin}
\end{equation}
The DM flux arriving at the Earth from Cen A can be derived from this rate as
\begin{equation}
    \frac{d\Phi_\oplus}{dE_\chi}
    =
    \frac{1}{d_L^2}
    \frac{dN_\chi}{dt dE_\chi d\Omega_\chi} ,
    \label{flux_spectrum_new_01}
\end{equation}
where $d_L = 3.7~\mathrm{Mpc}$ is the luminosity distance of Cen A.

\begin{table}[!t]
    \centering
    \begin{tabular}{lllll}
        \hline \hline 
        Parameter                                             & Symbol                                                     & Model $A$              & Model $A'$                & Model $B$ \\
        \hline 
        Luminosity Distance                                   & $d_L (\mathrm{Mpc})$                                       & 3.7                    & 3.7                       & 3.7                  \\
        Black hole mass                                       & $M_\mathrm{BH} (M_\odot)$                                  & $5  \times 10^7$        & $5  \times 10^7$           & $5  \times 10^7$ \\
        Boost factor of the blob                              & $\Gamma_B$                                                 & 3                      & 3                         & 3 \\
        Viewing angle of the jet      & $\theta_{\mathrm{LOS}}$                                    & $\mathbf{42^{\circ}}$  & $\mathbf{42^{\circ}}$     & $\mathbf{15^{\circ}}$ \\
        Minimal boost factor of electron      & $\gamma'_{\mathrm{e}, \min }$                              & $1$                    & $1$                       & $1$ \\
        Break boost factor of electron        & $\gamma'_{b }$                                             & $1.2  \times 10^3$      & $1.2  \times 10^3$         & $1.2  \times 10^3$ \\
        Maximal boost factor of electron      & $\gamma'_{\mathrm{e}, \max }$                              & $10^6$                 & $10^6$                    & $10^6$ \\
        Electron spectral power-law index       & $\alpha_{1}$                                               & 1.7                    & 1.7                       & 1.7 \\
        Electron spectral power-law index       & $\alpha_{2}$                                               & 4                      & 4                         & 4   \\
        Radius of the blob                    & $R'_B (\mathrm{cm})$                                       & $2.2 \times 10^{15}$    & $2.2 \times 10^{15}$       & $2.2 \times 10^{15}$  \\
        Electron luminosity                   & $L'_{\mathrm{e}}(\mathrm{erg} / \mathrm{s})$               & $10^{42}$              & $10^{42}$                 & $10^{42}$   \\
        \hline  
        DM profile power-law index                   & $\gamma_{\mathrm{sp}}$                                     & $\frac{7}{3}$          & $\frac{7}{3}$             & $\frac{7}{3}$        \\
        DM annihilation cross section & $\left\langle \sigma v_\mathrm{rel} \right\rangle  (\mathrm{cm}^3 / \mathrm{s})$ & $\mathbf{0}$           & $\mathbf{10^{-28}}$       & $\mathbf{0}$ \\
        Black hole lifetime                                   & $t_{\mathrm{BH}} (\mathrm{yr})$                            & $10^{9}$               & $10^{9}$                  & $10^{9}$ \\
        \hline \hline 
    \end{tabular}
    \caption{Parameters of three models for boosted DM from Centaurus A, including electron distribution parameters in the jet and DM density profile parameters.
        The differences between models are marked in bold.
    }
    \label{tab:parameters}
\end{table}

In this work, the parameters for the jet of Cen A used to calculate the boosted DM flux are adopted from Ref.~\cite{Banik:2020ffo} and summarized in Tab.~\ref{tab:parameters} as model $A$.
For DM profile, we take the DM annihilation cross section to be zero and the black hole lifetime to be $10^9$ years in model $A$ following Ref.~\cite{Wang:2021jic}.
We also introduce the model $A'$ that adjusts the annihilation cross section to be $10^{-28}~\text{cm}^3/\text{s}$ based on model $A$.
Additionally, the uncertainty of the viewing angle between the jet axis and our line-of-sight to Cen A is quite large, measured to be $\theta_{\mathrm{LOS}} \sim 12^\circ \mathrm{-} 45^\circ$ by the TANAMI program~\cite{Muller:2014wja}.
To illustrate the variations in boosted DM flux for different viewing angles, we define model $B$ based on model $A$ by replacing the value of $\theta_{\mathrm{LOS}}$ from $42^\circ$ to $15^\circ$.

\begin{figure}[t]
    \centering
    \includegraphics[width=0.7\textwidth]{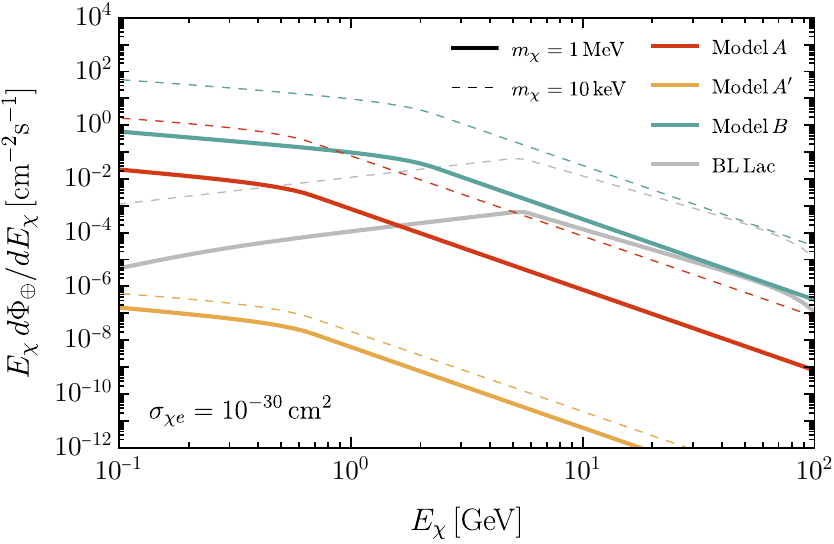}
    \caption{The spectrum of boosted DM flux from Centaurus A with DM-electron scattering cross section $\sigma_{\chi e} = 10^{-30} \, \mathrm{cm}^2$. 
    The DM mass is $1\, \mathrm{MeV}$ for the solid curves and $10\, \mathrm{keV}$ for the dashed curves.
    The red, yellow, and green curves correspond to model $A$, $A'$, and $B$, respectively. 
    }
    \label{fig:flux}
\end{figure}

In Fig.~\ref{fig:flux}, we show the spectra of boosted DM flux from Cen A calculated with the three models and different DM particle masses.
We can see that the DM flux is larger in model $B$ than that in model $A$, since the viewing angle $\theta_{\mathrm{LOS}}$ in model $B$ is smaller.
DM flux is significantly suppressed in model $A'$, since the DM number density in the spike is much lower in this model due to DM annihilation. 
Comparing the two cases with different DM particle masses in Fig.~\ref{fig:flux}, the flux of the lighter one is larger due to the enhanced DM number density in the spike.

In addition, we can compare the boosted DM flux from Cen A to that from the blazar BL Lacertae (BL Lac), whose jet points towards the Earth.
The electron luminosities, the masses of the central supermassive black holes, and hence the number densities of DM particles in the spike are similar in these two AGNs~\cite{Boettcher:2013wxa,Banik:2020ffo}.
Therefore, the total numbers of boosted DM particles from Cen A and BL Lac are comparable.
On one hand, our scenario experiences a suppression from the fraction of boosted DM propagating towards the Earth, which is around two orders of magnitude. 
On the other hand, Cen A is around $10^2$ times closer than BL Lac, leading to a four-order-of-magnitude enhancement in DM flux.
Consequently, the boosted DM flux from Cen A is expected to be about two orders of magnitude larger than that from BL Lac.

Note that we do not include the averaging factor in Eq.~\eqref{eq:effective_electron_distribution} for boosted DM flux from BL Lac to reproduce results in Ref.~\cite{Granelli:2022ysi}.
The parameters for the DM spike of BL Lac is set to be the same as model $A$.
Due to our more conservative treatment about the averaging of the electron density for Cen A, the DM flux from model $A$ is about one order of magnitude higher than that from BL Lac, as shown in Fig.~\ref{fig:flux}.

%%%%%%%%%%%%%%%%%%%%%%%%%%%%%%%%%%%%%%%%%%%%%%%%%%%%%%%%%%%%%%%%
\section{Dark Matter Attenuation}\label{sec:attenuation}
%%%%%%%%%%%%%%%%%%%%%%%%%%%%%%%%%%%%%%%%%%%%%%%%%%%%%%%%%%%%%%%%

The boosted DM from Cen A can be detected in the Super-Kamiokande (Super-K) experiment.
The Super-K experiment is located at a depth of $\sim 1$~km underground~\cite{Super-Kamiokande:2002weg}.
Before arriving at the underground detector, the boosted DM particles may lose energy and deflect due to elastic scattering off the electrons within the Earth, resulting in a distorted underground DM flux.
This phenomenon is known as the Earth attenuation effect.
For sufficiently large cross sections, this effect can be strong enough to completely shield the DM signal from the underground detector.

The Earth attenuation effect can be estimated by solving the one-dimensional energy loss equation~\cite{Starkman:1990nj,Bringmann:2018cvk,Ema:2018bih}, which is appropriate when particles travel in straight lines.
Other analytic methods taking into account the deflection of DM particles based on single scattering~\cite{Kavanagh:2016pyr} and the summation of multiple scatterings~\cite{Cappiello:2023hza} are also developed.
A more sophisticated method is the numerical Monte Carlo simulation~\cite{Collar:1992qc,Mahdawi:2017cxz,Emken:2017qmp,Chen:2021ifo,CDEX:2021cll}, which simulates the trajectory of DM particles and statistically reconstructs the underground DM flux.
This method accounts for the three-dimensional propagation and scattering of DM particles, as well as the spherical geometry of the Earth.
In this work, the underground boosted DM flux is simulated using the \texttt{DarkProp} code~\cite{DarkProp}.
This computational tool has been used to analyze the attenuation of relativistic cosmic ray boosted DM for both DM-nucleus~\cite{Xia:2021vbz} and DM-electron~\cite{Xia:2022tid} scatterings, and to study the attenuation of non-relativistic halo DM~\cite{Qiao:2023pbw}.

The lower boundary of the exclusion region on the DM-electron scattering cross section in this work turns out to be small enough that the Earth is almost transparent to DM particles.
So the simulation of the Earth attenuation focuses on determining the upper boundary of the cross section exclusion region, where DM particles lose significant energies that they cannot trigger the detector even if the zenith direction of the detector points to Cen A.
In this case the core and mantle structures of the Earth is not important.
Therefore, we adopt a simplified homogeneous Earth model with chemical composition and density the same as the outer crust, in which the number density of electrons is $n_e \approx 8 \times 10^{23}~\text{cm}^{-3}$.
Since the energy of DM particles considered in this work must be higher than the threshold of electron recoil energy in Super-K detector, which is $0.1$~GeV, the binding effect of the electrons within the atoms is neglected.
In our simulation, DM particles are injected from the same direction with energies sampled from the surface flux, and the simulation of each trajectory is terminated either when the particle leaves the Earth or its energy becomes lower than $0.1$~GeV.
After simulation, the underground flux is reconstructed as a global average by collecting all the crossing events at the spherical surface at the depth of 1 km, as described in Ref.~\cite{Xia:2021vbz}.

\begin{figure}[t]
    \centering
    \includegraphics[width=0.7\textwidth]{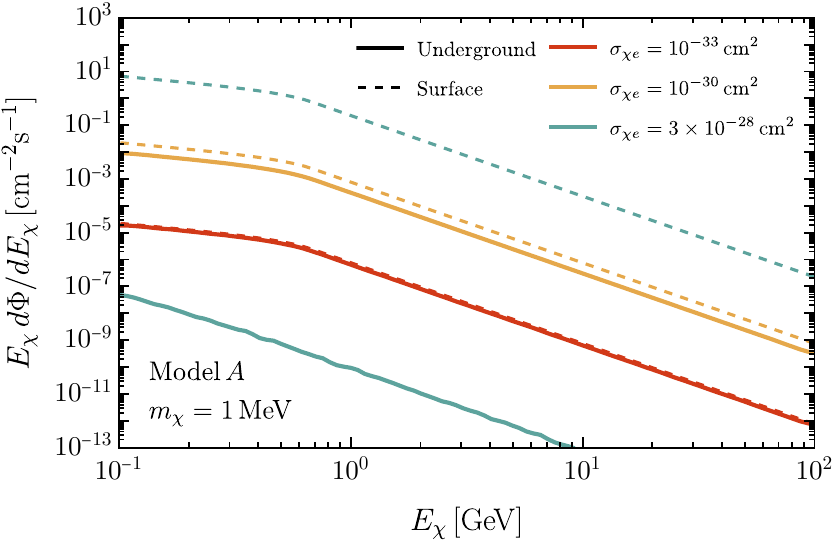}
    \caption{
    The underground boosted DM fluxes corresponding to model $A$ with $m_\chi = 1$~MeV for cross sections $\sigma_{\chi e} = 10^{-33}$, $10^{-30}$, and $3\times 10^{-28}~\text{cm}^2$.
    The surface fluxes without attenuation are also shown in dashed lines for comparison.
    }
    \label{fig:underground_flux}
\end{figure}

The reconstructed underground boosted DM fluxes corresponding to model $A$ are shown in Fig.~\ref{fig:underground_flux} with DM particle mass $m_\chi = 1$~MeV for different cross sections.
For the small cross section $\sigma_{\chi e} = 10^{-33}~\text{cm}^2$, the underground flux nearly coincides with the surface flux.
This can be elucidated by estimating the mean-free-path of DM particle traveling in the Earth, which is $\ell \equiv (n_e \sigma_{\chi e})^{-1} \approx 1.25 \times 10^4$~km for $\sigma_{\chi e} = 10^{-33}~\text{cm}^2$.
This length is comparable to the diameter of the Earth, so the Earth attenuation effect is negligible for $\sigma_{\chi e} \lesssim 10^{-33}~\text{cm}^2$.
For the larger cross section $\sigma_{\chi e} = 10^{-30}~\text{cm}^2$, $\ell$ reduces to $12.5$~km.
DM particles from the front side of the detector can easily cross the 1 km depth overburden, while those coming from the other side of the Earth are blocked.
Therefore, the Earth blocks half of the flux resulting in a factor 2 reduction compared to the surface flux, as shown in Fig.~\ref{fig:underground_flux}.
For $\sigma_{\chi e} = 3\times 10^{-28}~\text{cm}^2$, the mean-free-path is only about 41.7 m.
Thus, the 1 km overburden causes strong attenuation and the underground flux is reduced more than 10 orders of magnitude.
At such large cross sections, although the surface flux is proportional to $\sigma_{\chi e}$, the underground flux decreases exponentially with $\sigma_{\chi e}$ increasing.

%%%%%%%%%%%%%%%%%%%%%%%%%%%%%%%%%%%%%%%%%%%%%%%%%%%%%%%%%%%%%%%%
\section{Detection and Constraints}\label{sec:detection}
%%%%%%%%%%%%%%%%%%%%%%%%%%%%%%%%%%%%%%%%%%%%%%%%%%%%%%%%%%%%%%%%

After passing through the Earth, DM particles can scatter on target electrons in Super-K detectors and produce detectable Cherenkov light, which can be used to constrain the region of DM-electron scattering cross section. Since the signal energy $(> 0.1~\rm GeV)$ under consideration is much higher than the energy of target electrons, we assume reasonably that target electrons in the detector are at rest.
The differential event rate per unit target mass can be written as
\begin{equation}
	\frac{d\Gamma}{dE_e}
	=\mathcal{N}_e
	\int_{E_{\chi}^{\min}}dE_{\chi}
	\frac{\sigma_{\chi e}}{\Delta E^\text{max}_e} 
	\frac{d \Phi_{\oplus}}{dE_{\chi} } ,
	\label{eq:rate}
\end{equation}
where $\mathcal{N}_e$ is the number of electrons per unit target mass, taken as $\mathcal{N}_e \approx 3.3 \times 10^{26}~\rm{kg^{-1}}$ for water Cherenkov detectors, and $E_\chi^{\min}$ and $\Delta E_e^{\max}$ are given by the expressions of $E_e^{\min}$ and $\Delta E_\chi^{\max}$ in Eq.~\eqref{eq:production_rate} with $\chi$ and $e$ interchanged, respectively.

The Super-K Collaboration has reported a search for boosted DM using $161.9~\rm{kt\cdot yr}$ data in Ref.~\cite{Super-Kamiokande:2017dch}, which gives data for three energy bins, including the coordinates of event points, predicted backgrounds and signal detection efficiencies.
In this work, the analysis is based on the first energy bin in the range of $0.1\sim 1.33~\mathrm{GeV}$ of this data set.
Since the boosted DM particles come from the point source of Cen A and the scattering is almost forward, the direction of recoil electrons are nearly the same.
We therefore define a proper searching cone around the Cen A that includes most of the recoil directions.
The half-opening angle of the cone is determined by making the probability of the electron recoil direction in the cone equal 0.95, for the minimal incoming DM energy reproducing electron recoil energy that exceeds the threshold $0.1~\mathrm{GeV}$,
\begin{equation}
    \left.
    \int_{\cos{\delta}}^1
    du^e_{\rm sc}
    \frac{1}{\sigma_{\chi e}}
    \frac{d\sigma_{\chi e}}{d u^e_{\rm sc}}
    \right|_{E_\chi = E_\chi^\mathrm{th}}
    \simeq 0.95 \,,
    \quad \text{with} \quad
    E_\chi^\mathrm{th} \equiv E_\chi^{\min} ( E_e = 0.1~\mathrm{GeV} )\,,
    \label{eq:delta}
\end{equation}
where $\frac{d\sigma_{\chi e}}{d u^e_{\rm sc}}$ has the same form as Eq.~\eqref{dif_cross_section} with $\chi$ and $e$ interchanged, and $u^e_{\rm sc}$ is the scattering angle of the electron.
The numerical solution to Eq.~\eqref{eq:delta} gives $\delta\approx 23.7^\circ$.

We use the standard Poisson method to constrain the DM cross section for different DM masses. Given an expectation value of $\lambda=s+b$ events with $s$ the theoretical prediction and $b$ the expected background, the probability of observing $N_{\mathrm{obs}}$ events is given by the Poisson distribution
$P(N_{\mathrm{obs}}|\lambda)$.
The 95$\%$ confidence level (C.L.) upper limit $\lambda_p$ is determined by requiring $P(N<N_{\rm obs}|\lambda_p) = 0.05$.
Then the exclusion region of the cross section $\sigma_{\chi e}$ is obtained from $\lambda < \lambda_p$.

The theoretical prediction $s$ is given by
\begin{equation}
   s= 0.95\times \epsilon \times E_{\rm X}\int dE_e \frac{d\Gamma}{dE_e},
\end{equation}
where the coefficient 0.95 comes from the fact that the search cone contains 95\% of the total events, as implied by Eq.~\eqref{eq:delta}, $\epsilon$ is the signal efficiency taken as 0.93~\cite{Super-Kamiokande:2017dch}, $E_{\rm X}=161.9~\rm{kt\cdot yr}$ is the exposure of the Super-K data, and the integral is performed from 0.1~GeV to 1.33~GeV following the first bin of the data. For the background event number $b$, we use an isotropic distribution from the Super-K Monte Carlo simulation~\cite{Super-Kamiokande:2017dch}, which implies $b\approx168.4$ for $\delta=23.7^\circ$. The observed events $N_\mathrm{obs}$ within the search cone around Cen A direction, $(201.4^\circ, -43.0^\circ)$ in equatorial coordinate system, can be selected from the data, which gives $N_\mathrm{obs}=177$.

\begin{figure}[t]
    \centering
    \includegraphics[width=0.7\textwidth]{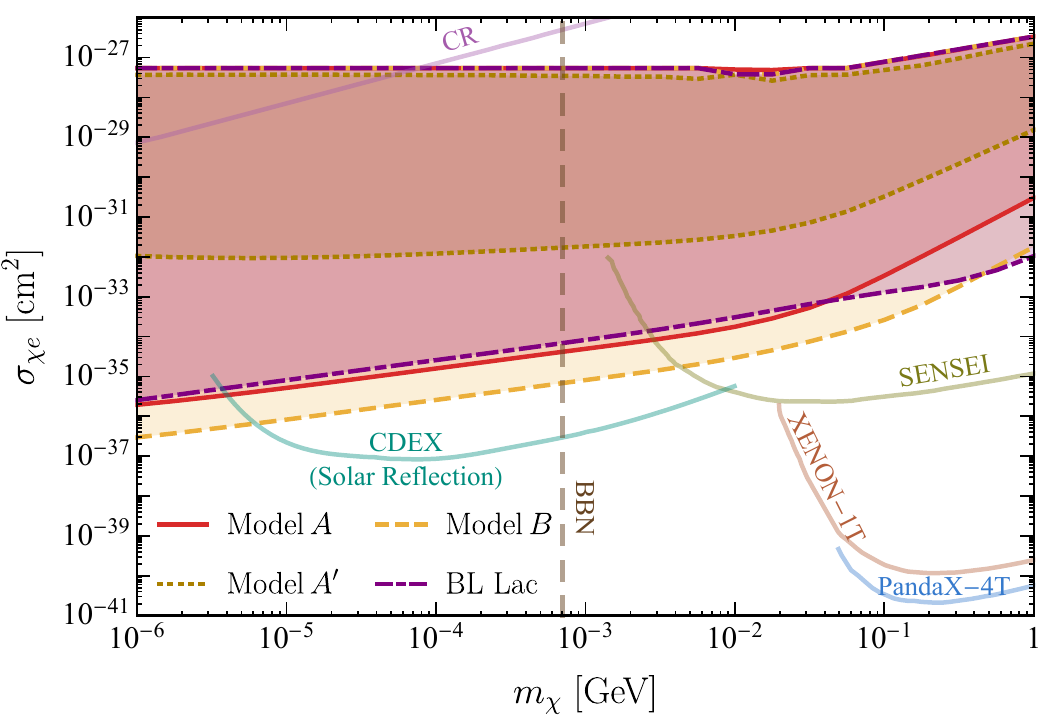}
    \caption{Exclusion regions in the $(m_\chi,\sigma_{\chi e})$ plane at 95\% C.L. derived from the first bin of Super-K data for model $A$, model $A'$, model $B$ of boosted DM from Cen A and the blazar BL Lac (shadow area). A selection of other constraints from XENON-1T~\cite{XENON:2019gfn}, PandaX-4T~\cite{PandaX:2022xqx}, SENSEI~\cite{SENSEI:2023zdf}, cosmic ray~\cite{Cappiello:2018hsu}, and CDEX with solar reflection~\cite{Zhang:2023xwv} are shown for comparison. The vertical brown dashed line stands for the BBN constraint on thermalized Dirac fermionic DM from Ref.~\cite{Sabti:2019mhn}.}
    \label{fig:constrain}
\end{figure}

In Fig.~\ref{fig:constrain}, we show the exclusion regions at 95\% C.L. calculated from the three models studied in this work, as well as that from the blazar BL Lac for comparison.
For model $A$, the lower boundary of the exclusion region is derived to be $5\times 10^{-35}~\text{cm}^2$ for $m_\chi = 10^{-3}$~GeV, and the constraint is more stringent for lighter DM.
If the DM annihilation channel is open, as in model $A'$, the constraint becomes weaker, since the DM number density in the spike is lower.
In model $B$, the viewing angle of Cen A is taken as $\theta_{\rm LOS} = 15^\circ$. 
The constraint can be one order of magnitude stronger than that from model $A$ with $\theta_{\rm LOS} = 42^\circ$.
Comparing model $A$ to BL Lac, we have estimated that the boosted DM flux from the former is about 10 times higher than the latter.
However, the constraint on the cross section from Cen A is not as low as around $\sqrt{10}$ times lower than that for BL Lac, as expected from that the number of DM events is proportional to $\sigma_{\chi e}^2$.
This is because the number of observed events in the searching cone around Cen A is larger than that of BL Lac in Super-K data, and the limit for Cen A is moderately weakened.
The upper boundary of the exclusion region due to the Earth attenuation is derived to be $\sim 5\times 10^{-28}~\mathrm{cm}^2$.
Larger cross sections may be probed by astrophysical observations such as cosmic ray energy loss~\cite{Cappiello:2018hsu}.
Other constraints from selected DM direct detection experiments, SENSEI~\cite{SENSEI:2023zdf}, XENON-1T~\cite{XENON:2019gfn}, PandaX-4T~\cite{PandaX:2022xqx}, and CDEX with solar reflection~\cite{Zhang:2023xwv}, as well as big bang nucleosynthesis (BBN) constraints on Dirac fermionic DM~\cite{Sabti:2019mhn} are also shown in Fig.~\ref{fig:constrain}.

%%%%%%%%%%%%%%%%%%%%%%%%%%%%%%%%%%%%%%%%%%%%%%%%%%%%%%%%%%%%%%%%

\section{Conclusion}\label{sec:summary}
%%%%%%%%%%%%%%%%%%%%%%%%%%%%%%%%%%%%%%%%%%%%%%%%%%%%%%%%%%%%%%%%

There are various high energy processes in astrophysics, and dark matter can be boosted in these environments. 
The recoil energy of boosted dark matter can exceed the detection threshold in underground experiments.
We study the production and detection of boosted dark matter from the jet of Centaurus A, the closest active galactic nucleus.
Since the jet of Centaurus A is not directed towards the Earth, the angle of the momentum of boosted dark matter against the jet axis should be quite large so that they can propagate to the Earth. 
Although most of the boosted dark matte particles travel along the jet, our research shows that the fraction of boosted dark matter with large oblique angles is still sizable. 
Due to the proximity of Centaurus A, compared to that from blazars, boosted dark matter flux from Centaurus A is enhanced by around two orders of magnitude. 
These boosted dark matter particles are detectable in the Super-Kamiokande experiment, which results in an exclusion limit on sub-GeV dark matter-electron scattering cross section as low as $\sim 10^{-36} \, \mathrm{cm}^{2}$.
Considering the Earth attenuation effect, the upper boundary of the exclusion region is set to be $\sim 5\times 10^{-28}~\mathrm{cm}^2$.

%%%%%%%%%%%%%%%%%%%%%%%%%%%%%%%%%%%%%%%%%%%%%%%%%%%%%%%%%%%%%%%%
\acknowledgments
%%%%%%%%%%%%%%%%%%%%%%%%%%%%%%%%%%%%%%%%%%%%%%%%%%%%%%%%%%%%%%%%

We thank Jin-Wei Wang and Shang-Ming Chen for helpful discussions.
This work is supported by the National Natural Science Foundation of China (NSFC) No.~12247141 and No.~12247148, 
by the Natural Science Foundation of Shandong Province under the grants No. ZR2023QA149, 
by the Project of Shandong Province Higher Educational Science and Technology Program under grants No. 2022KJ271,
and by NSFC No.12375101, 12090060 and 12090064.
The Monte Carlo simulations in this paper were run on the Siyuan-1 cluster supported by the Center for High Performance Computing at Shanghai Jiao Tong University.

\bibliographystyle{JHEP}
\bibliography{CenA-BDM}

\end{document}